\documentclass[12pt,preprint]{aastex}

\usepackage{subfigure}

\begin{document}

\title{Toward a standard Gamma Ray Burst: tight correlations between the prompt and the afterglow plateau phase emission}

\author{Maria Giovanna Dainotti\altaffilmark{1}, Micha{\l} Ostrowski\altaffilmark{1}, Richard Willingale\altaffilmark{2}}

\altaffiltext{1}{Obserwatorium Astronomiczne, Uniwersytet Jagiello\'nski, ul. Orla 171, 31-501 Krak{\'o}w, Poland: dainotti@oa.uj.edu.pl, mio@oa.uj.edu.pl}

\altaffiltext{2}{Department of Physics \& Astronomy, University of Leicester, Road Leicester LE1 7RH, United Kingdom: rw@star.le.ac.uk}


\begin{abstract}

To reveal and understand astrophysical processes responsible for the Gamma Ray Burst (GRB) phenomenon, it is crucial to discover and understand relations between their observational properties. The presented study is performed in the GRB rest frames and it uses a sample of 62 long GRBs from our sample of 77 Swift GRBs with known redshifts. Following the earlier analysis of the afterglow {\it characteristic luminosity $L^*_a$ -- break time $T^*_a$} correlation for a sample of long GRBs \citep{Dainotti2010} we extend it to correlations between the afterglow and the prompt emission GRB physical parameters. We reveal a tight physical scaling between the mentioned afterglow luminosity $ L^*_a$ and the prompt emission {\it mean} luminosity  $<L^*_p>_{45} \equiv E_{iso}/T^*_{45}$. The distribution, with the Spearman correlation coefficient reaching $0.95$ for the data subsample with most regular light curves, can be fitted with approximately $L^*_a \propto {<L^*_p>_{45}}^{0.7}$. We also analyzed correlations of $L^*_a$ with several other prompt emission parameters, including the isotropic energy $E_{iso}$, the peak energy in the $\nu F_{\nu}$ spectrum, $E_{peak}$, and the variability parameter, $V$, defined by \cite{N000}.  As a result, we reveal significant correlations also between these quantities, with an exception of the variability parameter. The main result of the present study is the discovery that the highest correlated GRB subsample in the \citet{Dainotti2010} afterglow analysis, for the GRBs with canonical X\,-\,ray light curves, leads also to the highest {\it prompt-afterglow} correlations and such events can be considered to form a sample of standard GRBs for astrophysics and cosmology.

\end{abstract}


\section{Introduction}

To better understand processes responsible for GRBs and possibly to create a new GRB-based cosmological standard candle, one should discover relations between their observational properties. Finding out universal properties or correlations for GRBs afterglows could be revealed by looking for strict relations among their observables. But, GRBs seem to be everything but standard candles, with their energetics spanning over eight orders of magnitude. However, the revealed correlations of $E_{iso}$\,-\,$E_{peak}$ \citep{amati09}, $E_\gamma$\,-\,$E_{peak}$ \citep{G04,Ghirlanda06}, $L$\,-\,$E_{peak}$ \citep{S03,Yonekotu04}, $L$\,-\,$V$ \citep{FRR00,R01} and other luminosity indicators \citep{FRR00,N000,liza05,LZ06} proposed allowed for expecting a quick progress in the field. The problem of large data scatter in the considered luminosity relations \citep{Butler2009,Yu09} and a possible impact of detector thresholds on cosmological standard candles \citep{Shahmoradi09} is also a debated issue \citep{Cabrera2007}. The underlying problem of the scatter in all the correlations is that it is larger than the spread expected from the z dependence alone. GRBs can be seen from a large fraction of the visible Universe, up to z=8.26. The luminosity spread due to, exclusively, its luminosity distance squared dependence gives for the limiting redshifts a factor of $D_L^2(8.26)/D_L^2(0.085)= 4.7 \times 10^{4}$ while the actual spread in luminosity is 8 orders of magnitude $10^{46}$ to $10^{54}$ ergs/s. It is not clear what is responsible for such a large dynamic range.

Among various attempts, \citet{DCC} have proposed a way to standardize GRBs as distance indicator with the discovery of $\log L^*_a$--$\log T^*_a$ (`LT') anti-correlation, where $L^*_a \equiv L^*_X(T_a)$\footnote{Note a change of notation with respect to our previous papers, where we used the symbol $L^*_X$ -- without an index `a' -- to indicate $L^*_X(T^*_a)$.}  is an isotropic X-ray luminosity in the time $T^*_a$, the transition time separating the plateau and the power-law decay afterglow phases and, henceforth, we use the index `$*$' to indicate quantities measured in the GRB rest frame.  We have presented \citep{Dainotti2010} an analysis revealing that the long GRBs with canonical light curves are much more tightly LT correlated as compared to the full sample of long GRBs. One may note that an analogous LT relation was derived phenomenologically by \citet{Ghisellini2009} and \citet{Yamazaki09} and that the LT correlation is also a useful test for the models of \citet{Cannizzo09} and \citet{Dall'Osso}. 

GRBs have been traditionally classified as {\it short/hard} ($T_{90}<2 s$) and {\it long/soft} ($T_{90}>2s$). The parameter of $T_{90}$, which is defined as the time interval during which the background-subtracted cumulative counts increase from $5\%$ to $95\%$, is usually used to denote the time duration of the GRB \citep{Kouveliotou:1993yx}. Some recent studies (see, e.g., Norris \& Bonnell 2006) have revealed the existence of an intermediate GRB class (IC), what requires a revision of the above simple scheme. In our analysis we consider only long GRBs, but on the plots with GRB distributions we also show the IC class events demonstrating remarkable differences as compared to the long bursts.  

In this Letter we study correlations between the {\it afterglow phase} luminosity parameter $L^*_a$ and the energetics and mean luminosity of the {\it prompt emission}. We demonstrate existence of significant correlations among the afterglow plateau and the prompt emission phases, which reach maximum for the {\it Swift} `canonical' light curve GRBs, the ones well fitted by a simple analytical expression proposed by \citet{W07}. The revealed high correlations indicate the expected  physical coupling between the GRB prompt and afterglow energetics, which is quite tight for the regular afterglow lightcurve GRBs (called in \citet{Dainotti2010} the {\it upper envelope}).
We also find that the prompt-afterglow correlations are more significant if one uses the prompt emission mean luminosity instead of the energy $E_{iso}$. This work reveals an important fact: any search for physical relations between GRB properties should involve selection of well constrained physical GRB subsamples. Usage of all available data introduces into analysis the events with highly scattered intrinsic physical properties, what smooths out possible correlations, and may lead to systematic shifts of the fitted relations, e.g. \citet{Dainotti2010}.  It is likely that a substantial fraction of the observed large scatter is introduced because we are observing different classes of GRBs with different progenitors and/or in different physical conditions. Identifying such subclasses may be the real challenge. Separating short and long GRBs is too simplistic. Below, we demonstrate that a particular class of canonical GRBs exists within the full sample of long GRBs. 
In the paper we use CGS units: [erg] for energy, [erg/s] for luminosity and [s] for time. All quantities used for correlation analysis are computed in the GRB rest frames (we indicate such quantities using a superscript *, $E_{iso}$ is in GRB rest frame from its definition).

\section{Data selection and analysis}

We can estimate the characteristic luminosity of a burst using different characteristic times, $T_{45}$, $T_{90}$ and $T_p$, where $T_{45}$
is the time spanned by the brightest 45 per cent of the total counts above the background \citep{R01} and $T_p$ is the fitted transition  time in which the exponential decay in the prompt phase changes to a power law decay. Here we define  $<L^*_p>_{45}\equiv E_{iso}/T^*_{45}$,  $<L^*_{p}>_{90} \equiv E_{iso}/T^*_{90}$ and $<L^*_{p}>_{Tp}\equiv E_{iso}/T^*_{p}$ and we have analyzed correlations between logarithms of the prompt emission parameters $E_{iso}$, $<L^*_p>_{45}$ , $<L^*_{p}>_{90}$, $<L^*_{p}>_{Tp}$, $ E_{peak}$, $V$ and the parameters $L^*_a$ and $T^*_a$ characterizing the afterglow light curve. 

The GRB sample used in the analysis is composed of all afterglows with known redshifts detected by Swift from January 2005 up to April 2009 \citep{Dainotti2010}, with the light curves possessing the early XRT data, enabling fitting by the \citet{W07} phenomenological model . The redshifts $z$ are taken from the Greiner's web site http://www.mpe.mpg.de/$\sim$jcg/grb.html~. We have compared these redshifts with the values reported by \citet{Butler2007,Butler2010} and we find that they agree well apart from two cases of GRB 050801 and 060814, but Butler (private communication, February 2010) suggested that we should use the Greiner redshifts for those two cases. For original references providing the redshift data see \citep{Butler2007,Butler2010}. The $E_{iso}$, $E_{peak}$, $T_{90}$ and $T_{45}$ values are listed in \citet{Butler2007,Butler2010}, values of the variability parameter $V$ are taken from \citet{S09}. The fitted values of $T_p$ used the for determination of $L^*_a$ by \citet{Dainotti2010} are given on the online data table http://www.oa.uj.edu.pl/M.Dainotti/GRB2011. Derivations of $T^*_a$ and $L^*_a$ for each afterglow follow \citet{DCC}, \citet{Dainotti2010} and \citet{W07}:

\begin{equation}
L^*_a=\frac{4\pi D_L^2(z) F_X(T_a) }{(1+z)^{1-\beta_{a}}} \qquad ,
\label{eq: finalLx}
\end{equation}

\noindent
where $D_L(z)$ is the source luminosity distance, $\beta_a$ is the X-ray spectral index of the emission at $T_a$, and the flux $F_X(t)$ derived for the time $t = T_a$ is obtained using the temporal evolution of the light curve \citep{W07} as 

\begin{equation}
F_X(T_a) =F_a \exp{\left ( - \frac{T_p}{T_a} \right )}  \qquad .
\label{eq: fluxafterglow}
\end{equation}

\noindent
We have computed the luminosities assuming the spectrum could be fitted with a simple power-law, see \citet{Dainotti2010} and \citet{Evans2009}.  Below, the fitted power-law relation between the analyzed quantities `$X$' and `$Y$' is $\log X =  \log b + a \cdot \log Y$ on the logarithmic plane, where the constants $a$ and $b$ are determined using the \citet{Dago05} method.  

In the original derivation of $E_{iso}$ by \citet{Amati02} the total radiated energy of a GRB is obtained in a fixed energy range, by integrating the best-fit model in the range 1--$10^4$ keV, for a given source luminosity distance. D$_L$ is derived assuming a flat Friedman-Robertson--Walker cosmological model with H$_0$ = 71 km s$^{-1}$ Mpc$^{-1}$, $\Omega$$_m$ = 0.3, $\Omega$$_\Lambda$ = 0.7. If $N(E$, $\alpha$, $E_0$,$\beta$, $A)$ is the best fit \citep{band93} model to the time-integrated and redshift-corrected spectrum of the GRB, $E_{iso}$ is given by

\begin{equation}
E_{iso} = \frac{ 4 \pi D^2_L}{(1 + z)^2} \cdot \int_{1}^{10^4}{E N(E,\alpha,E_0,\beta,A) dE}  \qquad .
\label{eq: eiso}
\end{equation}

\noindent
To use these $E_{iso}$ values taken from the literature  the luminosities $L^*_a$ have  been computed here using the given above cosmological parameters, see the online table, contrary to \cite{Dainotti2010}, where slightly different values were used.  

Our analyzed sample of 77 GRBs with the redshift range $0.08-8.26$ includes 66 long GRBs afterglows and 11 GRBs whose nature is debated, the claimed intermediate class (IC) between long and short GRBs. IC class is described by \cite{Norris2006} as an apparent (sub)class of bursts with a short initial pulse followed by an extended low-intensity emission phase. Our long GRB sample includes also 8 X-Ray Flashes (XRFs)(060108, 051016B, 050315, 050319 \citep{Gendre2007}, 050401, 050416A, 060512, 080330 \citep{Sakamoto2008}). To constrain the study to physically homogeneous samples we have analyzed the subsamples of 66 long GRBs (including XRFs) and of 11 IC ones separately, following the approach adopted in \citet{Dainotti2010}. From a homogeneous sample of long GRBs we extract subsamples of GRBs with improving  \cite{W07} canonical light curve fit quality. As a measure of the fit accuracy we use the respective logarithmic errors bars, $\sigma_{L^*_a}$ and $\sigma_{T^*_a}$, for the $L^*_a$ and $T^*_a$ parameters characterizing the GRB afterglow, to formally define a fit error parameter \citep{Dainotti2010}:

\begin{equation}
u \equiv \sqrt{\sigma_{L^*_a}^2 + \sigma_{T^*_a}^2}  \qquad .
\label{eq: uparameter}
\end{equation}

\noindent
In the study the limiting long GRB subsamples are: the largest one consisting of 62 long GRBs with $u \leq 4$, hereafter called `U4', and the previously called the {\it upper envelope} subsample, consisting of 8 GRBs with smallest afterglow fit errors, $u \leq 0.095$, hereafter called `U0.095'. We also analyze selected intermediate subsamples with the maximum $u$ values decreasing from 4.0 to 0.095, in attempt to reveal systematic variations of the studied correlations. This choice follows our previous paper, \citep{Dainotti2010}, and the discussion of systematics issues is presented in \citet{Dainotti2011} 
 
\section{`Prompt-afterglow' correlations}
\begin{figure}
\centering
\includegraphics[width=1.0\hsize,angle=0,clip]{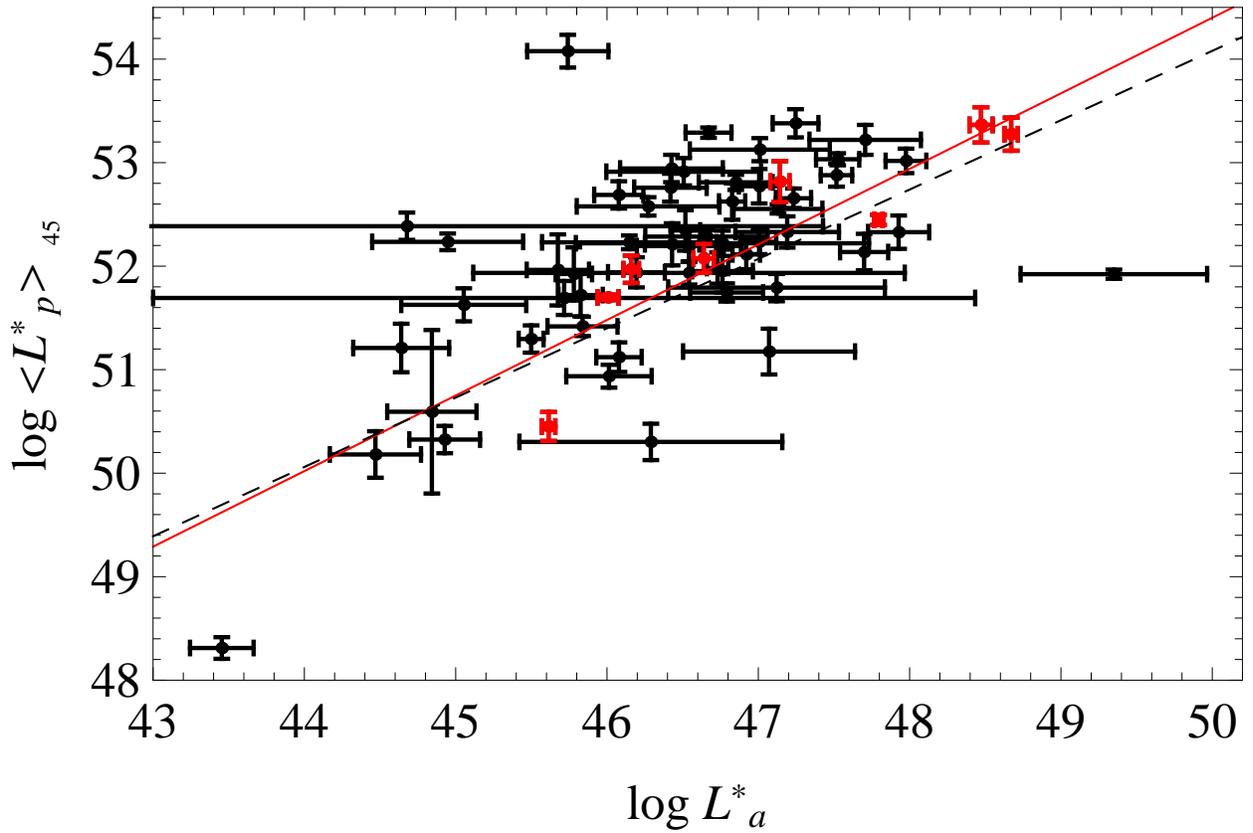}
\caption{$L^*_a$ versus $<L^*_{p}>_{45}$ distribution for the U4 sample (all points), with the fitted correlation dashed line in black. The red line is fitted to the 8 lowest error (red) points of the U0.095 subsample. 
\label{fig1}}
\end{figure}

\begin{figure}
\centering
\includegraphics[width=0.7\hsize,angle=270,clip]{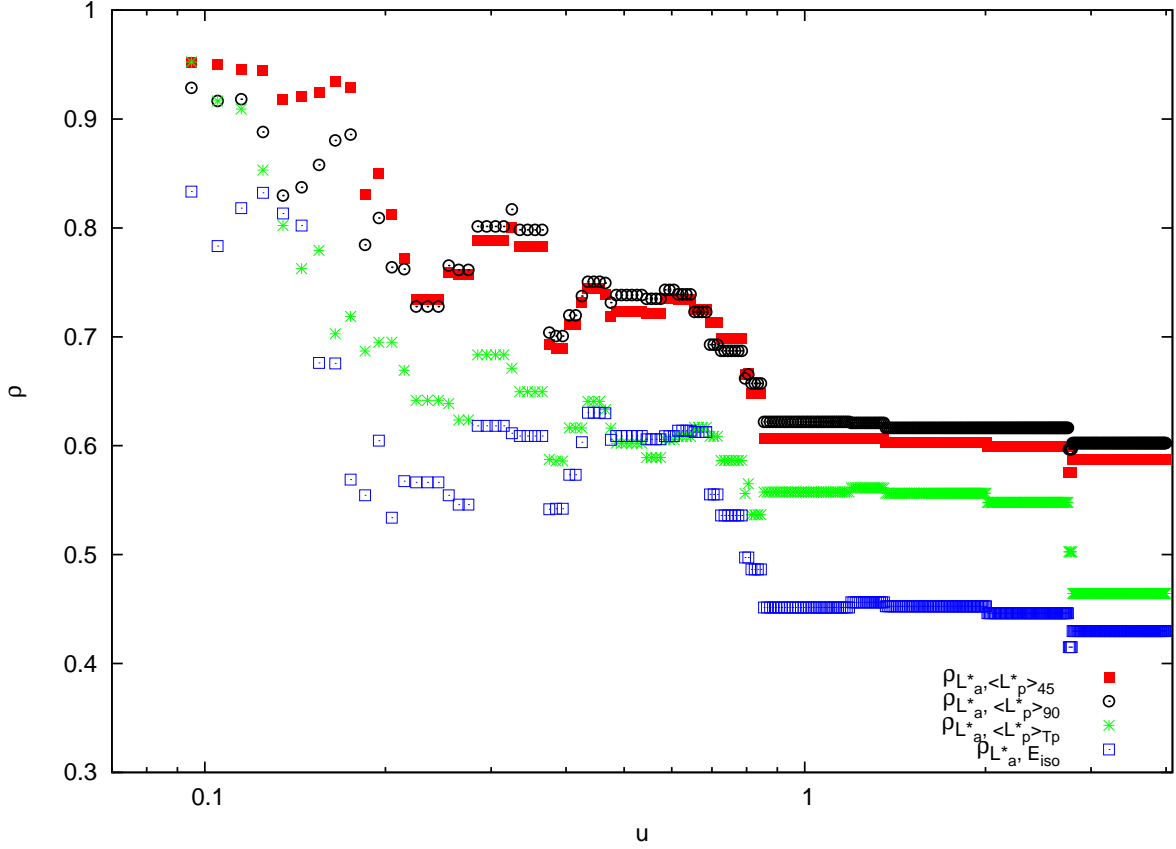}
\caption{Correlation coefficients $\rho$ for the distributions $\log L^*_a - \log <L^*_{p}>_{45}$ (red squares), $\log L^*_a - \log <L^*_{p}>_{90}$ (black circles),  $\log L^*_a - \log <L^*_{p}>_{Tp}$ (green asterixes) and $\log L^*_a - \log E_{iso}$ (blue squares) for the long GRB subsamples with the varying maximum error parameter $u$. 
\label{fig2}}  
\end{figure}

The derived $\log L^*_a$--$\log <L^*_{p}>_{45}$ distribution is presented for the U4 subsample of 62 long GRBs on Fig. \ref{fig1}, where, also, the U0.095 subsample of 8 GRBs with the most regular afterglow light curves is indicated. The distribution illustrates a significant correlation of the considered luminosities, with the Spearman correlation coefficient\footnote{A non-parametric measure of statistical dependence between two variables \citep{Spearman}.}, $\rho$, equal 0.64 for U4, but growing with rejecting high error points from the distribution (Fig. \ref{fig2}), to reach the value of 0.98 for U0.095 sample ($12 \%$ of the long GRBs). The other distributions considered in this study, involving $E_{iso}$, $<L^*_{p}>_{90}$, $<L^*_{p}>_{Tp}$ instead of $<L^*_{p}>_{45}$  also show significant correlations, with the lowest $u$ events forming in all cases tightly correlated subsamples of the full distribution (Fig. \ref{fig2}). The resulting correlation coefficients and the respective probabilities, $P=P(\rho \leq \rho_{pearson})$ \citep{Bevington}, generated by chance in a random distribution, and the parameters (a, b) of the fitted lines are given in Table \ref{Table1}.  

\begin{table}[t]
\caption{Correlation coefficients $\rho$, the respective correlation fit line parameters (a, $\log b$), and the correlation coefficient $\rho_{pearson}$ with the respective random occurrence probability $P$, for the considered `prompt-afterglow' and `prompt-prompt' distributions.}\label{Table1}
\begin{center}
\begin{tabular}{c|l c|c c}
\hline
 & \multicolumn{2}{c|}{U4} & \multicolumn{2}{c}{U0095} \\
\hline
Correlations & \multicolumn{2}{l|}{\hspace{20pt}$\rho$ \hspace{60pt}  a, $\log b$} & \multicolumn{2}{l}{\hspace{20pt}$\rho$ \hspace{60pt} a, $\log b$} \\
\cline{2-5} & \multicolumn{2}{l|}{P} & \multicolumn{2}{l|}{P} \\
\hline
$~$ & $~$ & $~$ & $~$ & $~$ \\

$L^*_a-<L^*_{p}>_{45}$ & 0.59 & $(0.67 _{-0.15}^{+0.14}, 20.58 _{-6.73}^{+6.66})$  & 0.95 & $(0.73 _{-0.11}^{+0.16}, 17.90_{-6.0}^{+5.29})$  \\
& 0.62 &  $7.7 \times 10^{-8}$  & 0.90& $2.3 \times 10^{-3}$ \\
$~$ & $~$ & $~$ & $~$ & $~$ \\
\hline
$~$ & $~$ & $~$ & $~$ & $~$ \\
$L^*_a-<L^*_{p}>_{90}$ & 0.60 & $(0.63 _{-0.16}^{+0.15}, 22.05_{-7.31}^{+7.14})$ & 0.93 &  $(0.84 _{-0.12}^{+0.11}, 11.86_{-3.44}^{+3.43})$  \\
& 0.62 & $7.7 \times 10^{-8}$ &0.94 & $2.7 \times 10^{-3}$\\
$~$ & $~$ & $~$ & $~$ & $~$ \\
\hline
$~$ & $~$ & $~$ & $~$ & $~$ \\
$L^*_a-<L^*_{p}>_{Tp}$ & 0.46 & $(0.73_{-0.14}^{+0.09}, 16.61_{-4.35}^{+4.35})$  & 0.95 & $(0.93_{-0.23}^{+0.20},7.70_{-3.46}^{+3.47})$\\
& 0.56 & $2.21 \times 10^{-6}$ &0.90 & $2.3 \times 10^{-3}$\\
$~$ & $~$ & $~$ & $~$ & $~$ \\
\hline  \hline
$~$ & $~$ & $~$ & $~$ & $~$ \\
$L^*_a-E_{iso}$ & 0.43 & $(0.52 _{-0.06}^{+0.07}, 28.03_{-2.97}^{+2.98})$  & 0.83 & $(0.49 _{-0.16}^{+0.21}, 29.82 _{-7.82}^{+7.11})$ \\
 &0.52 & $1.4 \times 10^{-5}$ & 0.75 & $3.2\times 10^{-2}$ \\
$~$ & $~$ & $~$ & $~$ & $~$ \\
\hline 
$~$ & $~$ & $~$ & $~$ & $~$ \\
$T^*_a-E_{iso}$ &  -0.19 & $(-0.49_{-0.08}^{+0.09},54.51_{-0.30}^{+0.37})$  &  -0.81 &$(-0.96_{-0.22}^{+0.21}, 54.67_{-0.69}^{+0.69})$\\
&-0.21 & $1.0 \times 10^{-1}$  & -0.69 & $5.8 \times 10^{-2}$\\
$~$ & $~$ & $~$ & $~$ & $~$ \\
\hline
$~$ & $~$ & $~$ & $~$ & $~$ \\
$L^*_a-E_{peak}$ & 0.54 & $(1.06_{-0.23}^{+0.53}, 43.88_{-1.00}^{+0.61})$  & $0.74$ & $(1.5_{-0.94}^{+0.65}, 43.10_{-2.26}^{+2.53})$\\
& 0.51& $2.2 \times 10^{-5}$ &0.80 & $1.7 \times 10^{-2}$\\
$~$ & $~$ & $~$ & $~$ & $~$ \\
\hline
$~$ & $~$ & $~$ & $~$ & $~$ \\
$T^*_a-E_{peak}$ &  -0.36 & $(-0.66_{-0.29}^{+0.20}, 4.96_{-0.80}^{+0.81})$ &  -0.74 &  $(-1.40_{-0.65}^{+0.66}, 7.04_{-1.77}^{+1.79})$\\
& -0.35& $5.2 \times 10^{-3}$ & -0.77 & $2.5 \times 10^{-2}$ \\
$~$ & $~$ & $~$ & $~$ & $~$ \\
\hline \hline
$~$ & $~$ & $~$ & $~$ & $~$ \\
$<L^*_{p}>_{45}-E^*_{peak}$ & 0.81 & $(1.14_{-0.25}^{+0.22}, 49.27_{-0.60}^{+0.61})$ &  0.76 &   $(1.45, _{-0.54}^{+0.26}, 48.48_{-1.04}^{+1.05})$\\
&0.67&  $2.6 \times 10^{-9}$ & 0.92& $1.2 \times 10^{-3}$\\

\hline
\end{tabular}
\end{center}
\end{table}

\begin{table}
\caption{Online Table: A data list for GRBs with known redshifts analysed in the paper: 62 long GRBs with $u<4$ (upper part of the table) and 11 IC GRBs (Lower part of the table). The U0.095 subsample is indicated by ash attached to their u values.}\label{OnlineTable}
\begin{center}
\begin{tabular}{ccccccc}
\hline
$Id_{GRB}$ & $z$ & $beta_a$ & $F*x$ & $log T*_{90}$ & $log T*_{45}$ & $log T*_p$\\
\hline
$~$ & $~$ & $~$ & $(erg/cm^2*s) $ & $(s)$ & $(s)$ & $(s)$\\
                	      	                                           
050315 &		1.95	&0.89 ± 0.04&	4.58E-12 ± 1.97E-12&		1.510 ± 0.014&		0.816 ± 0.018&	 	1.010 ± 0.037	\\	
050318&		1.44&	0.93 ± 0.18	&1.00E-08 ± 1.41E-08&		1.100 ± 0.001	&	0.157 ± 0.024 & -1.770 ± 0.132 \\
050319	&	3.24	&0.85 ± 0.02&	4.31E-12 ± 1.78E-12&		1.560 ± 0.005	&	0.452 ± 0.032	&  0.260 ± 0.099	\\	
050401	&	2.9	&	1.00 ± 0.04	 &3.87E-11 ± 1.33E-11 &		0.945 ± 0.006	&	0.125 ± 0.033 & -1.480 ± 0.316\\
050416A	&	0.65 &	0.99 ± 0.10	& 1.06E-11 ± 5.62E-12	&	0.245 ± 0.044 & -0.418 ± 0.041 &	 	0.042 ± 0.105		\\
050505	&	4.26	& 1.09 ± 0.04	& 4.93E-12 ± 3.84E-12	&	1.060 ± 0.007	&	0.260 ± 0.036	& 	0.137 ± 0.882\\	
050603	&	2.82 &	0.91 ± 0.10	& 1.10E-12 ± 6.64E-13	&	0.409 ± 0.026 & -0.378 ± 0.027 &	 -0.230 ± 0.199\\	
050730 &		3.97 &	0.52 ± 0.27 &	6.59E-11 ± 1.12E-11	&	1.080 ± 0.021 &		0.626 ± 0.021	& 	0.695 ± 0.052\\
050801 &		1.56	& 1.43 ± 0.30 &	4.23E-11 ± 1.62E-11 &		0.363 ± 0.036 & -0.408 ± 0.043 &	 	0.690 ± 0.197\\	

\end{tabular}
\end{center}
\end{table}

Fig. \ref{fig2} illustrates the trend in a few tested `prompt-afterglow' distributions  to increase the correlation coefficient with selecting the GRBs with more regular {\it afterglow} light curves, as measured by the error parameter $u$. The same trend was presented earlier by \cite{Dainotti2010} for the afterglow ($\log L^*_a$, $\log T^*_a$) distribution. On the figure, e.g., we have data derived for 62 long GRBs for $u=4$, 33 GRBs for $u=0.3$, 19 GRBs for $u=0.15$, 13 GRBs for $u=0.12$ and 8 GRBs left for the limiting $u=0.095$. The prompt emission parameters $E_{iso}$, $<L^*_{p}>_{90}$ and $<L^*_{p}>_{Tp}$ tested versus the afterglow luminosity $L^*_a$ show significant correlations (cf. Table \ref{Table1}), but one should note that the mean prompt emission luminosity, $<L^*_{p}>_{45}$, derived using the characteristic time scale $T_{45}$, provides the highest value of the Spearman correlation coefficient \footnote{When comparing the times $T_{90}$ and $T_{45}$ the better one to represent the GRB prompt emission energetics when we consider the canonical lightcurves at $u<0.25$ is the second one, as the first one is more dependent on the BAT detector sensitivity limit, see \citet{Willingale2010}.}. One may also note that the correlations involving the considered mean prompt emission luminosities are higher that the one involving the isotropic energy $E_{iso}$.

\begin{figure}
\includegraphics[width=0.5\hsize,angle=0,clip]{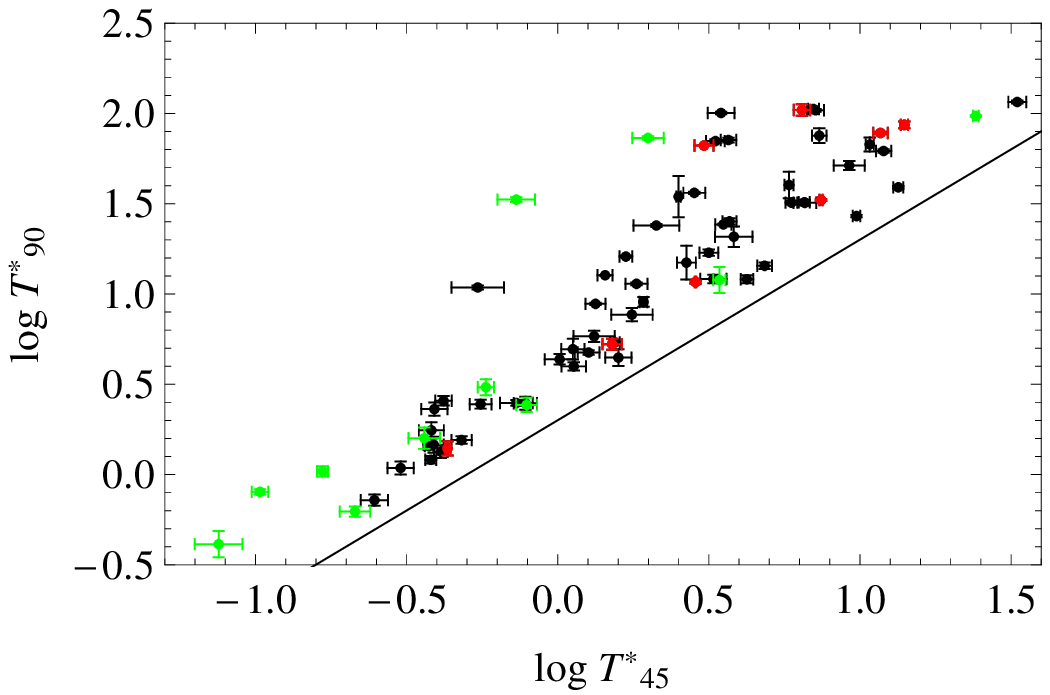}
\includegraphics[width=0.5\hsize,angle=0,clip]{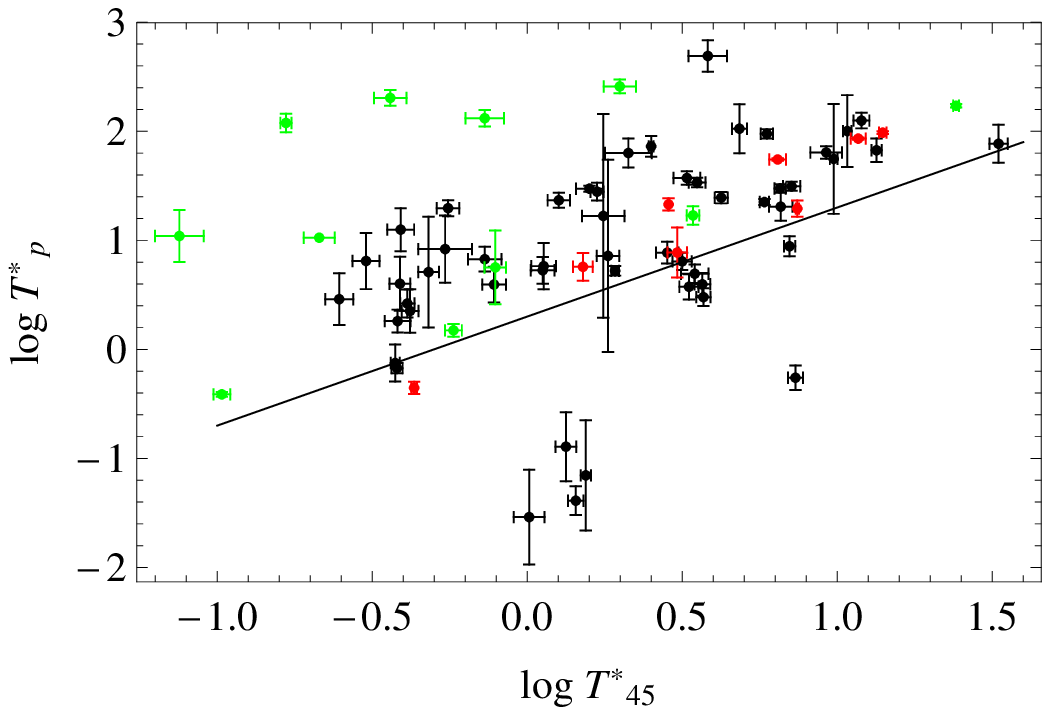}
\caption{Comparison of the characteristic time scales for the GRB prompt emission for all GRBs analyzed in this paper. The green points are the IC GRBs, the red ones are the long GRBs with $u\leq 0.095$ and the black ones are the other long GRBs with $u \leq 4$. Left panel: $\log T^*_{90}$ vs $\log T^*_{45}$ distribution.  Right panel: $\log T^*_p$ vs $\log T^*_{45}$ distribution. The reference lines are $ T^*_{90}=2*T^*_{45}$ and $T^*_{45}= T^*_{p}$ for the left and right panel respectively.  
\label{fig3}}  
\end{figure}

The GRB energy flux of the prompt emission phase is highly non-uniform, non-evenly distributed within the time $T_{90}$ or $T_p$, as compared to $T_{45}$ (Fig. \ref{fig3}). Thus selecting different characteristic time scales to derive the mean prompt luminosity is equivalent to considering different physical phases of the prompt emission variation.
$T_{45}$ puts greater emphasis on the peaks of the luminosity, while $T_{90}$ including periods when the emission is low or absent  puts therefore more weight on the total elapsed time of the activity period.
Our analysis suggests that the $T^*_{45}$ time scale better represents the prompt emission energetics than the other considered times, at least when we relate its to the afterglow luminosity and the more uniform lightcurves.

After the above comparison of the considered correlations we conclude that presence of tight correlations involving the prompt emission quantities for a small $u$ subsample, defined from the afterglow light curves only, proves that such sample forms a well defined physical class of standard GRBs with tight relations between their prompt emission and the afterglow light curve properties. Having such a tool to extract GRB events enables to reveal a number of strict relations between their observational parameters, partly hidden otherwise within large scattered samples involving all available events. In the considered standard GRBs the mechanism causing the prompt phase of the burst influences directly the afterglow plateau phase, as discussed, e.g., by  \citet{Shao2007,Liang2007,Troja2007,Ghisellini2007,Nava2007,Racusin2009} and \citet{Shen2009}.

To better understand how the afterglow plateau phase properties are related to the instantaneous or averaged physical parameters of the prompt emission, we have investigated the following additional distributions: $E_{iso}-T^*_a$, $E^*_{peak}-L^*_a$, $E^*_{peak}-T^*_a$, $V-L^*_a$ (see the central part of Table \ref{Table1}). For the $E^*_{peak}-L^*_a$ distribution we obtain significant correlation as one could expect from the known $E^*_{peak}-E^*_{iso}$ correlation \citep{amati09}: $\rho_{L^*_a, E^*_{peak}}=0.54$ for the U4 subsample, growing to $0.74$ for the U0.095. We have also found significant  $E_{peak}-<L^*_{p}>_{45}$ shown at the bottom of the table (for a similar correlation see \citet{Collazzi2008}).
Furthermore, since $L^*_a$ is anti-correlated with $T^*_a$ we derived the expected correlations involving the time scale $T^*_a$ for the distributions $T^*_a$ -- $E^*_{peak}$ and  $T^*_a$ -- $E^*_{iso}$. We note that the $\rho$ of these correlations involving the timescale $T_a$ are smaller than the ones which correlate the prompt energetic and  $L^*_a$. Instead, for $V-L^*_a$ we did not find any significant correlation or any clear trend for decreasing $u$ subsamples (cf \citet{Lloyd2002,RamirezRuizFenimore1999} for the $V-E_{peak}$ correlations and analogous relations) when these energies are transformed to the cosmological rest frame. 

If we compare the $a$ and $\log b$ values given in Table \ref{Table1} for the U4 and U0.095 samples we find a good agreement within the error bars. One should remember that the considered mean luminosities $<L>_T \equiv E_{iso}/T$ depend on the applied time scales $T$. 

No significant correlations between $L^*_a$ and the prompt emission quantities $<L^*_{p}>_{Tp}$, $<L^*_{p}>_{45}$ and $<L^*_{p}>_{90}$ exist for the $u<4$ subsample of IC GRB afterglows, including  050724, 051221A, 060614, 060502, 070810, 070809, 070714 \citep{Norris2009} and 060912A \citep{Levan2007}, but the number of events is too small to draw any firm conclusion from this fact.
Furthermore, for some of these bursts, the determination of the redshift is not so firm, therefore the conclusion of the lack of correlation on this sub-sample could change with an enlarged and more firm redshift sample.

\section{Final remarks}

In this Letter we present new significant correlations between the luminosity of the afterglow plateau phase, $L^*_a$, and numerous parameters of the prompt emission, including the mean luminosities and the integral energy derived for this emission. 
For the light curves which are smooth and well fitted by the \citet{W07} phenomenological model we find tight correlations in the analyzed distributions, showing that only GRBs with regular light curves exhibit strict physical scalings between their observed characteristics. Thus only such events can be considered to form the standard GRB sample, to be used for both GRB detailed physical model discussion and, possibly, to work out the GRB-related cosmological standard candle. A progress in both issues requires to increase an observed number of the canonical light curve GRBs, not by simply increasing the total number of GRBs with know redshifts. 

GRBs with the light curve non-uniformities exhibit weaker correlations of the plateau phase and the prompt emission energetics. No significant prompt-afterglow correlations were detected for the sample of IC GRBs, but the small number of registered events unable one to draw any firm conclusion about existence or not of such correlations for the canonical light curve shapes. From the pictures of the existing correlations it is clear that the inclusion of the GRB IC class do not strengthen the existing relations. Therefore, any future detailed study of the relations between various GRB properties should involve a separation of the IC GRBs from the long ones, to perform analysis using physically homogeneous sub-samples. 

Correlations between the physical properties of the prompt emission and the luminosity of the afterglow plateau reveals that mean (averaged in time) energetic properties of the prompt emission more directly influence the plateau phase as compared to $E_{iso}$, providing new constraints for the physical model of the GRB explosion mechanism, namely the plateau phase results related to the inner engine, as it has been already pointed out by \citet{Ghisellini2009}. In the analysis we show that the mean luminosities derived using the $T^*_{45}$ time scale better correlate to the afterglow luminosity than the ones applying the other considered time scales. 

\section{Acknowledgments}
This work made use of data supplied by the UK Swift Science Data Center at the University of Leicester. MGD and MO are grateful for support from the Polish Ministry of Science and Higher Education through the grant N N203 380336. MGD is also grateful for support from the Angelo Della Riccia Foundation.

\end{document}